\documentclass{amsart}
\usepackage{tikz}
\usetikzlibrary{decorations.markings}
\usetikzlibrary{decorations.pathreplacing}
\usepackage{amsmath}

\begin{document}

\title[A note on  ``Einstein's special
  relativity beyond the speed of light'']{A note on  ``Einstein's special
  relativity beyond the speed of light by James M. Hill and Barry J. Cox''}

\author{ Hajnal~Andr\'eka, Judit~X.~Madar\'asz, Istv\'an~N\'emeti, and
  Gergely~Sz\'ekely}

\address{Alfr\'ed R\'enyi Institute of Mathematics, Hungarian Academy
  of Sciences \\ Re\'altanoda utca 13-15, H-1053, Budapest, Hungary}


\begin{abstract}
We show that the transformations J.~M.~Hill and B.~J.~Cox introduce
between inertial observers moving faster than light with respect to
each other are consistent with Einstein's principle of relativity only
if the spacetime is 2 dimensional.
\end{abstract}


\maketitle

\section{Introduction}
J.~M.~Hill and B.~J.~Cox introduce the following two
transformations to extend Lorentz transformations for inertial
observers moving faster than light (FTL) with respect to each other,
see equations (3.16) and (3.18) in Hill \& Cox (2012):
\begin{equation}\label{HC1}
HC_1:\enskip
t=\frac{-T+vX/c^2}{\sqrt{v^2/c^2-1}},\enskip
x=\frac{-X+vT}{\sqrt{v^2/c^2-1}}, \enskip y=Y,\enskip  z=Z
\end{equation}
and
\begin{equation}\label{HC2}
HC_2:\enskip
t=\frac{T-vX/c^2}{\sqrt{v^2/c^2-1}},\enskip
x=\frac{X-vT}{\sqrt{v^2/c^2-1}}, \enskip y=Y,\enskip z=Z
\end{equation}
where $v$ is the superluminal relative speed of the two observers.

In the present paper, we show that Hill--Cox
transformations give a consistent extension of Einstein's special
theory of relativity {\it only} if the dimension $d$ of spacetime is
2 (i.e., there are 1 space and 1 time dimensions).

\section{Consistency of Hill--Cox transformations with the principle of relativity implies $d=2$}
Einstein originally formulated his principle as ``The laws by
  which the states of physical systems undergo change are not
  affected, whether these changes of state be referred to the one or
  the other of two systems of co-ordinates in uniform translatory
  motion.'' (see Einstein 1905). Clearly, this principle implies that
  inertial observers cannot be distinguished by physical experiments,
  e.g., by experiments based on sending out light signals.

We show that in the $HC_i$-transformed worldview, the light cone is
``flipped over'' so that its axis is the $x$-axis. Hence $x$ is a
unique direction in which the speed of light is smallest (namely, $c$);
in all other directions, either one cannot send out a light signal, or
the speed of the light signal is greater than $c$ (there are plenty of
these two types of direction).  From this, the FTL observer belonging
to the transformed reference frame can know/``observe'' that he is
moving FTL and $x$ is a distinguished unique direction in his
worldview. Further, there are directions in which no light signals can
be mirrored back and forth (like in Einstein's light clock) but not
all directions are such, while in the worldview of any slower than
light observer there are no such directions. This is quite a strong
  violation of Einstein's principle of relativity (e.g., because the
  space of each slower than light observer is isotropic while this is
  not so for the FTL observers).

To see what the Hill--Cox transformations do with the light cones if
$d>2$, first we show that they can be written as the composition of
a Lorentz transformation and a transformation exchanging the time
axis and a space axis if we use relativistic units (i.e., if
the speed $c$ of light is set to be 1)\footnote{The physical
  meaning of this choice is to fit the units of measuring time and
  distance together, i.e., measuring time in years and distance in
  light years, or measuring time in Planck time and distance in Planck
  length.}, otherwise a
scaling of time enters the picture, too.

Let $v,c$ be as in (\ref{HC1}),(\ref{HC2}). Let $L$ be the Lorentz
boost, in relativistic units, corresponding to velocity $c/v$, i.e., $L$ takes $(T,X,Y,Z)$ to
$(t,x,y,z)$  where

\begin{equation}\label{L}
L:\enskip t=\frac{T-(c/v)X}{\sqrt{1-(c/v)^2}},\enskip
  x=\frac{X-(c/v)T}{\sqrt{{1-(c/v)^2}}},\enskip y=Y,\enskip z=Z
\end{equation}

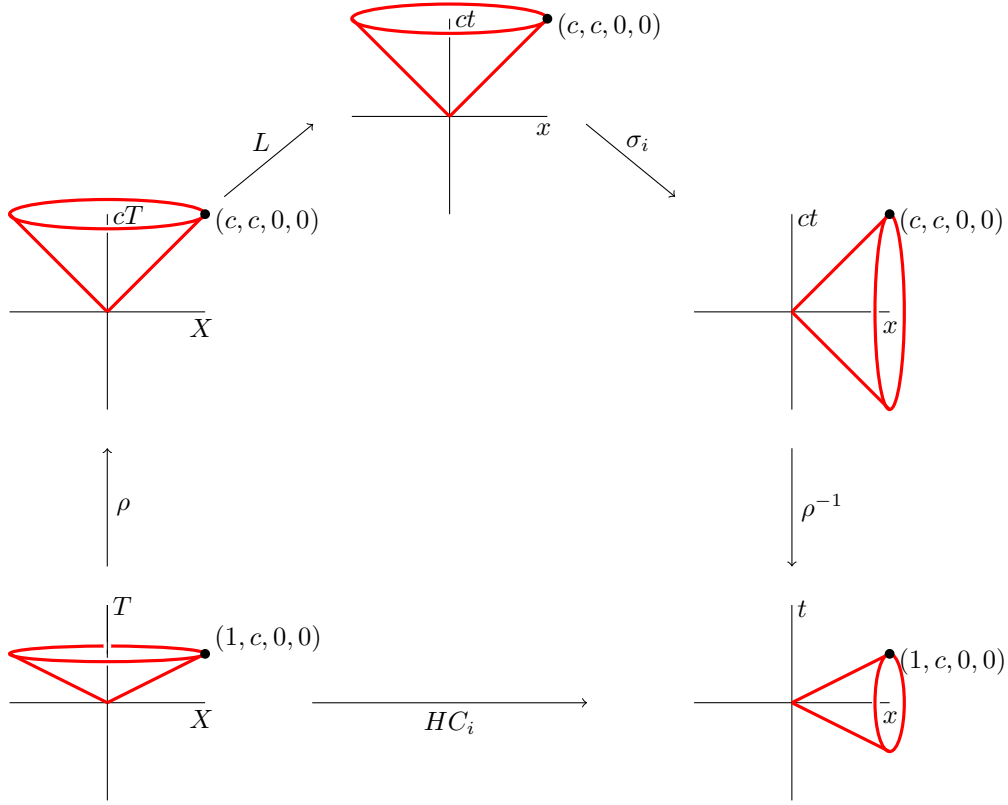
\begin{figure}
\begin{center}
\begin{tikzpicture}[scale=1.3]
\begin{scope}[xshift=-3.5cm]
\node (A1) at (0,-1.3){};
\node (A) at (1.1,1.1){};
\node (E) at (2,0){};
\draw (0,1) node[right,yshift=-.1mm,xshift=-0.5mm]{$cT$}-- (0,-1);
\draw (1,0) node[below,yshift=.3mm,xshift=-0.5mm] {$X$} -- (-1,0);
\fill[white] (0,0.85) circle (0.04);
\draw[red,very thick] (0,0) -- (-1,1);
\draw[red,very thick] (0,0) -- (1,1);
\draw[red,very thick] (0,1) ellipse (1 and .15);
\fill (1,1)  circle (0.05) node[right,yshift=-1mm] {$(c,c,0,0)$};
\end{scope}
\begin{scope}[xshift=-3.5cm,yshift=-4cm]
\node (A2) at (0,1.3){};
\node (G) at (2,0){};
\draw (0,1) node[right,yshift=-.1mm,xshift=-0.5mm]{$T$}-- (0,-1);
\draw (1,0) node[below,yshift=.3mm,xshift=-0.5mm] {$X$} -- (-1,0);
\fill[white] (0,0.42) circle (0.04);
\draw[red,very thick] (0,0) -- (-1,0.5);
\draw[red,very thick] (0,0) -- (1,0.5);
\draw[red,very thick] (0,0.5) ellipse (1 and .08);
\fill (1,0.5)  circle (0.05) node[right,yshift=2mm] {$(1,c,0,0)$};
\fill[white] (0,.58) circle (0.04);
\draw (0,0.5)--(0,1);
\end{scope}
\begin{scope}[yshift=2cm]
\node (B) at (-1.3,0) {};
\node (C) at (1.3,0) {};
\draw (0,1) node[right,yshift=-.0mm,xshift=-0.5mm]{$ct$}-- (0,-1);
\draw (1,0) node[below,yshift=0.3mm,xshift=-0.5mm] {$x$} -- (-1,0);
\fill[white] (0,0.85) circle (0.04);
\draw[red,very thick] (-1,1) -- (0,0);
\draw[red,very thick] (0,0) -- (1,1);
\draw[red,very thick] (0,1) ellipse (1 and .15);
\fill (1,1)  circle (0.05) node[right,yshift=-1mm] {$(c,c,0,0)$};
\end{scope}
\begin{scope}[xshift=3.5cm]
\node (B1) at (0,-1.3) {};
\node (D) at (-1.1,1.1) {};
\node (F) at (-2,0) {};
\draw (0,1)  node[right,yshift=-.5mm,xshift=-0.5mm]{$ct$}-- (0,-1);
\draw (1,0) node[below,yshift=0mm,xshift=0.1mm]{$x$} -- (0,0);
\fill[white] (0.85,0) circle (0.04);
\fill[white] (-0.85,0) circle (0.04);
\draw (0,0) -- (-1,0);
\draw[red,very thick] (0,0) -- (1,-1);
\draw[red,very thick] (0,0) -- (1,1);
\draw[red,very thick] (1,0) ellipse (.15 and 1);
\fill (1,1)  circle (0.05) node[right,yshift=-1mm] {$(c,c,0,0)$};
\end{scope}
\begin{scope}[xshift=3.5cm,yshift=-4cm]
\node (B2) at (0,1.3) {};
\node (H) at (-2,0) {};
\draw (0,1)  node[right,yshift=-.5mm,xshift=-0.5mm]{$t$}-- (0,-1);
\draw (1,0) node[below,yshift=0mm,xshift=0.1mm]{$x$} -- (0,0);
\draw (0,0) -- (-1,0);
\draw[red,very thick] (1,-0.5) -- (0,0);
\draw[red,very thick] (0,0) -- (1,0.5);
\fill[white] (0.86,0) circle (0.04);
\draw[red,very thick] (1,0) ellipse (0.15 and .5);
\fill (1,0.5)  circle (0.05) node[right,yshift=-1mm] {$(1,c,0,0)$};
\end{scope}
\draw[->] (A2) --node[right]{$\rho$} (A1);
\draw[->] (B1) --node[right]{$\rho^{-1}$} (B2);
\draw[->] (A) --node[above,xshift=-1mm]{$L$} (B);
\draw[->] (C) --node[above,xshift=1mm]{$\sigma_i$} (D);
\draw[->] (G) --node[below]{$HC_i$} (H);
\end{tikzpicture}
\end{center}
\caption{Decomposition of Hill--Cox transformations  \label{decomp}}
\end{figure}

Let us note that $c/v<1$ if $v>c$. Let
$\sigma_1$ be the transformation that interchanges the first two
coordinates, i.e., $\sigma_1$ takes $(T,X,Y,Z)$ to $(X,T,Y,Z)$, let
$\sigma_2$ be the transformation that takes $(T,X,Y,Z)$ to
$(-X,-T,Y,Z)$, and finally let $\rho$ be the transformation that
scales the time coordinate with $c$, i.e., $\rho$ takes $(T,X,Y,Z)$
to $(cT,X,Y,Z)$. Then a straightforward computation shows that

\begin{equation}
HC_i=\rho^{-1}\circ\sigma_i\circ L\circ\rho,
\end{equation}
see Figure~\ref{decomp} and the computation below. This decomposition
will make transparent what $HC_i$ do with the light cones. The key
point will be that transformations $\sigma_1$ and $\sigma_2$ radically
deform the light cones in $d$ dimensions if $d>2$, they only
preserve the light cones in $2$ dimensions.

\begin{eqnarray*}
  (\rho^{-1}\circ \sigma_1\circ L \circ \rho)(T,X,Y,Z) =\newline
  (\rho^{-1}\circ\sigma_1\circ L)(cT,X,Y,Z) =\\ (\rho^{-1}\circ
  \sigma_1)\left(\frac{cT-(c/v)X}{\sqrt{1-(c/v)^2}},\frac{X-(c/v)cT}{\sqrt{1-(c/v)^2}},Y,Z\right)
  =\\ \rho^{-1}\left(\frac{X-(c/v)cT}{\sqrt{1-(c/v)^2}},\frac{cT-(c/v)X}{\sqrt{1-(c/v)^2}},Y,Z\right)
  =\\ \left(\frac{X/c-(c/v)T}{\sqrt{1-(c/v)^2}},\frac{cT-(c/v)X}{\sqrt{1-(c/v)^2}},Y,Z\right)
  =\\ \left(\frac{(c/v)(vX/c^2-T)}{(c/v)\sqrt{(v/c)^2-1}},\frac{(c/v)(vT-X)}{(c/v)\sqrt{(v/c)^2-1}},Y,Z\right)=\\
\left(\frac{-T+vX/c^2}{\sqrt{v^2/c^2-1}},\frac{-X+vT}{\sqrt{v^2/c^2-1}},Y,Z\right)=\\ HC_1(T,X,Y,Z).
  \end{eqnarray*}
The computation for $HC_2$ is similar.

Now let us see what the $HC_i$-transformed light cones are like.
First we give a geometric visual proof for our original claim, and
then we supplement this proof with computations valid for any
$d>2$. The idea of our proof is depicted in Figure~\ref{decomp} and is
based on the ``step-by-step'' understanding of what Hill--Cox
transformations do with the light cones. Let us concentrate on the
light cone emanated from the origin.  In the original, non-transformed
worldview, this light cone is a regular cone with width $c$ (since the
speed of light is $c$ in each direction). Now, $\rho$ scales the time
axis such that this cone becomes a right-angle one, i.e., one with
width 1. Transformation $L$, being a simple Lorentz boost, acts
non-trivially only in plane $TX$ and takes this light cone to itself
(basic property of Lorentz transformations).  Then $\sigma_i$ flips
this cone over by exchanging coordinates $x$ and $t$. Finally,
$\rho^{-1}$ compresses the flipped-over cone in the $t$ direction so
that the ``height'' of this flipped-over cone is $c$ while its
``width'' remains 1.  Thus, after this final compression, the steepest
line of this flipped-over cone goes in the $x$ direction and the
corresponding speed is $c$; all spatial projections of the lines of
this cone enclose at most an $45^\circ$ angle with the $x$-axis. Hence
there are no light signals in any direction enclosing an angle greater
than $45^\circ$ with the $x$-axis and the speed of light is greater
than $c$ for every direction enclosing an angle (strictly) between
$0^\circ$ and $45^\circ$ with the $x$-axis. The speed of light is
infinite in the directions enclosing exactly $45^\circ$ angle with the
$x$-axis. By using some right-angle triangles, one can compute the
exact dependence of the speed $c(\alpha)$ of light going in a
direction that encloses an angle $\alpha$ with the $x$-axis:

\begin{equation}\label{calpha}
c(\alpha) = c\cdot\sqrt{\frac{1+\tan(\alpha)^2}{1-\tan(\alpha)^2}},\enskip \text{ where } 0\le \alpha< 45^\circ.
\end{equation}

Let us now give the calculations belonging to the above chain of
thoughts. We use $d=4$, but the computations we give are completely
analogous for any $d>2$.

The equation of the light cone (whose apex is the
origin) in the $(T,X,Y,Z)$ coordinate system is
\begin{equation}\label{lc}
(cT)^2=X^2+Y^2+Z^2
\end{equation}
We get the $HC_i$-image of this by successively applying $\rho, L,
\sigma_i, \rho^{-1}$ to this equation, and we get in both cases of
$i=1,2$
\begin{equation}\label{LC}
x^2=(ct)^2+y^2+z^2.
\end{equation}
Equation \eqref{LC} corresponds to the flipped-over light cone
depicted in the bottom right corner of Figure~\ref{decomp}.  Let
$\ell$ be any line going through the origin and orthogonal to the
time axis, see Figure~\ref{abra}. Then there are $A,B,C$ such that
$\ell$'s equation is
\begin{equation}\label{ell}
(0,As,Bs,Cs),\quad s\in \mathbb{R}.
\end{equation}
Thus the point $P$ of the light cone in direction $\ell$ and with
time coordinate 1 is
\begin{equation}
(1,As,Bs,Cs)\quad\mbox{with}\quad (As)^2=c^2+(Bs)^2+(Cs)^2
\end{equation}
from where we get
\begin{equation}
s=\frac{c}{\sqrt{A^2-B^2-C^2}},
\end{equation}
thus the speed of light in direction $\ell$ is
\begin{equation}\label{cell}
c(\ell)=\frac{c\cdot\sqrt{A^2+B^2+C^2}}{\sqrt{A^2-B^2-C^2}}.
\end{equation}
Let $\alpha$ denote the angle between $\ell$ and the $x$-axis, then
$\tan(\alpha)=\sqrt{B^2+C^2}/A$. Substituting this to (\ref{cell})
we get (\ref{calpha}).

\begin{figure}
\begin{center}
\begin{tikzpicture}[scale=6]
\coordinate (O) at (0,0);
\coordinate (E) at (1,0.5);
\coordinate (P) at (0.75,0.32);
\coordinate (F) at (0.75,0.5);
\coordinate (C) at (1.5,0.3);
\coordinate (B) at (0.5,-0.3);
\coordinate (A) at (0.75,-0.15);
\coordinate (Q) at (1,0);
\coordinate (D) at (1,0.8);
\draw[->,shorten <= -0.5cm,>=latex] (O) -- (0,0.75)  node[right,yshift=-.5mm,xshift=-0.5mm]{$t$};
\draw[->,shorten <= -0.5cm,>=latex] (O) node[left,yshift=-2mm]{$O$} -- (1.5,0)  node[below,yshift=0mm,xshift=0.1mm]{$x$};
\draw[red, thick] (0,0) -- (E);
\draw[red, thick] (0,0) -- (0.5,-0.3);
\draw[red, thick]  (B) .. controls (0.5,0) and (0.62,0.3) .. (E) .. controls (1.3,0.63) and (1.5,0.4).. (C) ;
\draw (0.4,0) arc (0:-45:.1); 
\draw (0.86,0) .. controls (0.83,-0.03) and (0.83,-0.05)  .. (0.89,-.065);
\fill (0.9,-0.03) circle (0.007); 
\draw (0.65,-.13) arc (180:90:.1);
\fill (0.71,-0.09) circle (0.007); 
\draw[shorten >= -2.5 cm, shorten <= -0.5cm, thick] (O) node[xshift=17mm,yshift=-1.5mm]{$\alpha$}-- (A) ;
\node at (1.2,-.3) {$\ell$};
\draw [decorate,decoration={brace,amplitude=3pt,mirror,raise=3pt},yshift=0pt] (O) -- (0.75,-0.16);
\draw [decorate,decoration={brace,amplitude=3pt,mirror,raise=5pt}] (A) -- (Q);
\draw [decorate,decoration={brace,amplitude=3pt,mirror,raise=4pt}] (A) -- (P);
\node at (0.82,0.09) {$1$}; 
\node[right] at (0.9,-0.12){$s\cdot\sqrt{B^2+C^2}$} ;
\node[below, rotate=-13] at (0.3,-0.1) {$s\cdot\sqrt{A^2+B^2+C^2}$};
\fill (1,0) circle (0.015);
\draw[ultra thick] (A) -- (Q);
\draw[thick] (A) -- (P);
\draw[ultra thick] (O) -- (Q);
\draw[red, very thick] (O) -- (P);
\draw (B) -- (C);
\draw[ultra thick] (O) -- (A);
\fill (P)  circle (0.015) node[right,yshift=3pt,xshift=2pt]{$P=(1,As,Bs,Cs)$};
\fill (O)  circle (0.015);
\fill (C)  circle (0.016);
\fill (B)  circle (0.016);
\fill (A)  circle (0.015);
\end{tikzpicture}
\caption{\label{abra} Illustration for the derivation of \eqref{calpha}}
\end{center}
\end{figure}

The fact that Hill--Cox transformations do not work if $d>2$ is not
surprising. It can be shown in a strictly axiomatic framework, with
using only a few assumptions of special relativity theory that
inertial observers cannot move faster than the speed of light if
$d>2$, see, e.g., Theorem 2.1 in Andr\'eka {\it et al.}  (2012). By
observers we mean reference frames as, e.g., the standard relativity
book d'Inverno (1992) does. So the difference between particles and
observers is that particles do not need to have worldviews (frames
of reference), hence dealing with particles does
not require dealing with worldview transformations.

For $d=2$, transformations $HC_1$ and $HC_2$ are perfectly
consistent with Einstein's special relativity. In this case,
exchanging time and space is the usual way for constructing models
satisfying the axioms of special relativity in which there are FTL
observers. This construction is investigated in section 2.4 in
Andr\'eka {\it et al.} (2002).

\section{Do we need FTL observers in a theory of FTL particles?}

The existence of particles moving with the speed of light (photons)
does not imply the existence of observers moving with the speed of
light. The same way, the existence of FTL particles does not imply
(logically) the existence of FTL observers. This fact suggests that
in order to elaborate a theory of superluminal particles, we do not
necessarily have to introduce superluminal observers.

Indeed, even though observers cannot move FTL if $d>2$, the
superluminal motion of particles is consistent with both the kinematics and the dynamics
of special relativity, see Andr\'eka {\it et al.} (2002), Sz\'ekely (2012), and Madar\'asz \& Sz\'ekely (2012).

\begin{figure}[h!tb]
\begin{center}
\tikzset{->-/.style={decoration={
  markings,
  mark=at position #1 with {\arrow{>}}},postaction={decorate}}}
\begin{tikzpicture}[scale=2]
\coordinate (o) at (0,0);
\draw[gray] (-2,-1) grid (2,2);
\draw[red, very thick] (1,-1) -- (o) -- (2,2);
\draw[red, very thick] (-1,-1) -- (o) -- (-2,2);
\draw[magenta, ultra thick] (-1.95,2) .. controls (-1.8,1.93) and (-1.2,1) .. (0,1) .. controls (1.2,1) and (1.8,1.93) .. (1.95,2);
\draw[magenta, ultra thick] (-1,0.04) .. controls (-1.1,1.2) and (-1.9,1.8) .. (-2,1.95);
\draw[magenta, ultra thick] (1,0.04) .. controls (1.1,1.2) and (1.9,1.8) .. (2,1.95);
\draw[magenta]  (1,0) circle (0.04);
\draw[magenta]  (-1,0) circle (0.04);
\draw[->-=.75,blue, ultra thick,>=stealth] (o) -- (1.2,0.8);
\draw[dashed, thin] (1.2,0) -- (1.2,0.8) -- (0,0.8) node[left] {$m_{k}(b)$};
\draw[dashed, thin] (1.2,0) -- (1.2,0.8) -- (0,0.8);
\fill (1.2,0.8) circle (0.04);
\fill (1.2,0) circle (0.04);
\fill (0,0.8) circle (0.04);
\fill (o) circle (0.04);
\draw [decorate,decoration={brace,amplitude=11pt},xshift=0pt,yshift=0pt]
(1,0) --node [yshift=-5mm,xshift=3mm] {$p_\infty$} (0,0);
\end{tikzpicture}
\end{center}
\caption{\label{p} Illustration for equations \eqref{inv} and \eqref{pinf}}
\end{figure}
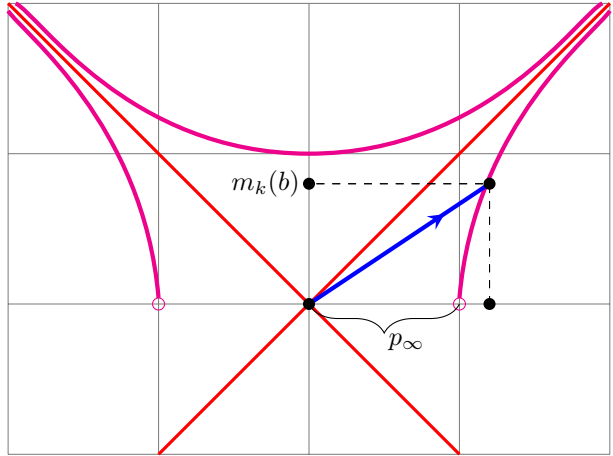

In Hill \& Cox (2012) it is shown that the relativistic mass
($m$) depends on the speed ($v$) of a superluminal particle and an
observer independent quantity $p_\infty$ as follows:
\begin{equation}\label{HCm}
m=\frac{p_\infty/c }{\sqrt{v^2/c^2-1}}.
\end{equation}

The dynamical results of Hill and Cox can also be proved to hold in a
strictly axiomatic framework without using FTL observers using very
few, simple assumptions. For example, their formula \eqref{HCm} can be
derived because a natural, consistent axiom system of special
relativistic particle dynamics containing Einstein's principle of
relativity implies that
\begin{equation}\label{inv}
m_k(b)\sqrt{\left|1-v^2_k(b)\right|}=m_h(b)\sqrt{\left|1-v^2_h(b)\right|},
\end{equation}
where $m_k(b)$ and $m_h(b)$ are the relativistic masses and $v_k(b)$
and $v_h(b)$ are the speeds of a (possibly FTL) particle $b$ with
respect to (ordinary slower than light) inertial observers $k$ and
$h$. This is done in Madar\'asz \& Sz\'ekely (2012), relying on
Andr\'eka {\it et al} (2008). We get formula \eqref{HCm} in
  relativistic units by introducing observer independent quantity for
FTL particle $b$ as
\begin{equation}\label{pinf}
p_\infty(b):=m_k(b)\sqrt{v^2_k(b)-1}.
\end{equation}

\section{Acknowledgment}
{This work is supported by the Hungarian Scientific Research
Fund for basic research grants No.~T81188 and No.~PD84093.}

\end{document}